\documentclass[twocolumn,preprintnumbers,floats,prd,amssymb,floatfix,nofootinbib,balancelastpage,superscriptaddress,amsmath]{revtex4-1}

\pdfoutput=1

\usepackage[utf8]{inputenc}
\usepackage{amsfonts}
\usepackage{amsmath}
\usepackage{amssymb}
\usepackage{amstext}
\usepackage{amsbsy}
\usepackage{bbm}
\usepackage{graphicx}
\usepackage{color}
\usepackage{xspace}
\usepackage{braket}
\usepackage{ulem}
\usepackage{cancel}
\usepackage[section]{placeins}
\usepackage{afterpage}
\usepackage{float}
\usepackage{slashed}
\usepackage{multirow,rotating}
\usepackage{orcidlink}

\newcommand{\lam}{\lambda}

\begin{document}

\title{The decay $B\to K+\nu+\bar{\nu}$ at Belle II and a massless bino \\ in R-parity-violating supersymmetry}

\author{Zeren Simon Wang\,\orcidlink{0000-0002-1483-6314}}
\email{wzs@mx.nthu.edu.tw}
\affiliation{School of Physics, Hefei University of Technology, Hefei 230601, China}

\author{Herbert K.~Dreiner\,\orcidlink{0000-0003-4963-1340}}
\email{dreiner@uni--bonn.de}
\affiliation{Physikalisches Institut der Universit\"at Bonn, Bethe Center for Theoretical Physics, \\ Nu{\ss}allee 12, 53115 Bonn, Germany}

\author{Julian Y. G\"unther\,\orcidlink{0000-0003-3233-4753}}
\email{guenther@physik.uni-bonn.de}
\affiliation{Physikalisches Institut der Universit\"at Bonn, Bethe Center for Theoretical Physics, \\ Nu{\ss}allee 12, 53115 Bonn, Germany}

\begin{abstract}
    Recently, \texttt{Belle II} announced evidence for the decay $B^+\to K^+ +\nu +\bar{\nu}$ at the $3.5\,\sigma$ significance, and measured the corresponding decay branching ratio to be $2.7\,\sigma$ above the Standard Model prediction.
    Here, we provide a theoretical explanation based on a massless bino in R-parity-violating (RPV) supersymmetry.
    With a single non-vanishing RPV coupling $\lambda'_{i23/i32}$, $\lambda'_{i13/i33}$, or $\lambda'_{i12/i22}$, where $i=1,2,3$, the decay $B^+\to K^+ +\overset{\scriptscriptstyle(-)}{\nu_i}+ \tilde{\chi}^0_1$ can be induced, which would lead to the same signature as $B^+\to K^+ +\nu +\bar{\nu}$. 
    Taking into account both theoretical and experimental uncertainties, we derive the regions of the parameter space spanned by the RPV couplings and sfermion masses that could accommodate the new measurement at various significance levels and obey the existing bounds simultaneously.
\end{abstract}

\vskip10mm

\maketitle
\flushbottom

\section{Introduction}\label{sec:intro}
Recently, \texttt{Belle~II} has presented the first evidence for the rare decay $B^+ \to K^+ +\nu+ \bar{\nu}$ with a measured branching 
ratio~\cite{Belle-II:2023esi}:
\begin{eqnarray}
    \text{BR}^{\text{exp}}(B^+ \to K^+ +\nu +\bar{\nu}) = (2.3 \pm 0.7) \times 10^{-5}.
    \label{eq:BelleII_result}
\end{eqnarray}
The significance is about $3.5\,\sigma$.
The Standard Model (SM) prediction~\cite{Parrott:2022zte} (see also Ref.~\cite{Becirevic:2023aov})
\begin{eqnarray}
    \text{BR}^{\text{SM}}(B^+\to K^+ +\nu+ \bar{\nu})=(5.6\pm 0.4)\times 10^{-6},
    \label{eq:SM_prediction}
\end{eqnarray}
is lower by about $2.7\,\sigma$.
This experimental enhancement in the branching ratio has since been interpreted from various theoretical perspectives~\cite{Asadi:2023ucx,Athron:2023hmz,Bause:2023mfe,Allwicher:2023xba,Felkl:2023ayn,He:2023bnk,Chen:2023wpb,Berezhnoy:2023rxx,Alonso-Alvarez:2023mgc,DaRold:2023hmx,Altmannshofer:2023hkn,Fridell:2023ssf,Ho:2024cwk,Hou:2024vyw,Chen:2024cll,Bolton:2024egx,Rosauro-Alcaraz:2024mvx,Buras:2024ewl}. 
In this short paper, we consider the possibility that this 
discrepancy is confirmed and hardened by upcoming data. We 
propose that it is due to an additional $B$-decay resulting 
from a supersymmetric model with broken R-parity 
(RPV-SUSY)~\cite{Dreiner:1991pe, Allanach:2003eb, 
Barbier:2004ez}. 

In supersymmetry in general, the spectrum of elementary 
particles is extended by a scalar Higgs doublet, and
then a doubling of the entire spectrum.
Each spin-1/2 particle (quark, lepton) now has a spin-0 scalar partner 
with identical internal quantum numbers (squark, 
slepton). Similarly the spin-1 gauge bosons have 
spin-1/2 gaugino partners, and the spin-0 Higgs bosons 
have spin-1/2 higgsino partners. The partners of the 
neutral gauge bosons $W^0$ (SU(2)$_{\text{L}}$) and $B$ 
(U(1)$_{\text{Y}}$), the wino and bino, as well as the 
two neutral higgsinos mix after electroweak symmetry 
breaking. The four resulting mass eigenstates are 
called neutralinos. In a series of papers we have shown 
that the lightest of these four neutralinos, denoted 
$\tilde\chi^0_1$ can indeed be very light, or even 
massless. Such a massless neutralino is consistent with all 
laboratory~\cite{Dreiner:1991pe, Choudhury:1999tn, Dreiner:2009er,Dreiner:2009ic,Gogoladze:2002xp,Dreiner:2022swd}, as well as astrophysical and cosmological~\cite{Dreiner:2003wh, Dreiner:2011fp, Dreiner:2013tja} constraints.
An overview is presented in Ref.~\cite{Dreiner:2009ic}.
There it was also shown, that a very light neutralino with mass $m_{\tilde\chi^0_1}<\mathcal{O}(1\,\text{GeV})$ is necessarily dominantly bino, \textit{i.e.}~it has the  quantum numbers $(\mathbf{1},\mathbf{1},0)$ under the SM gauge group $SU(3)_{\text C}\times SU(2)_{\text{L}}\times U(1)_{\text{Y}}$, just like a right-handed 
neutrino~\cite{Dercks:2018wum, Dreiner:2023gir}.

In supersymmetric models with broken R-parity, the superpotential beyond the
MSSM contains among others the lepton-number-violating terms
\begin{equation}
W_{\cancel{L}} =\lambda'_{ijk} L_iQ_j\bar D_k,
\label{eq:rpv-superpotential}
\end{equation}
where $L,Q,\bar D$ denote the lepton doublet, quark doublet, and down-like quark singlet chiral superfields, respectively, and $\lambda'_{ijk}$ are dimensionless Yukawa couplings with $i,j,k\in \{1,2,3\}$ being generation indices.
We follow the notation of Ref.~\cite{Allanach:2003eb}.

The operators in Eq.~\eqref{eq:rpv-superpotential} can modify known decay rates of mesons to SM final states.
Comparing with experiment, these lead to an extended set of upper bounds on the couplings $\lambda'_{ijk}$ as a function of the involved scalar fermion mass; see for example Refs.~\cite{Barger:1989rk, Allanach:1999ic, Barbier:2004ez}.
Below in Sect.~\ref{sec:model}, we present the bounds relevant to the couplings of our analysis.
With a light neutralino, the operators can also mediate the new 2-body decay modes via virtual sfermions~\cite{Choudhury:1999tn,Dedes:2001zia, Dreiner:2009er, deVries:2015mfw, Dercks:2018wum, Dercks:2018eua, Dreiner:2020qbi, Dreiner:2022swd}
\begin{eqnarray}
    M^0\to \tilde\chi^0_1 + \nu \,,\quad M^\pm \to \tilde\chi^0_1+\ell^\pm\,,
\end{eqnarray}
for neutral ($M^0$) or charged ($M^\pm$) mesons, as well as the new decays~\cite{Dreiner:2009er}
\begin{equation}
    M^0\to \tilde\chi^0_1 + \tilde\chi^0_1\,.
\end{equation}
They can also result in new three-body decay modes~\cite{Adhikari:1994iv, Dreiner:2009er}
\begin{equation}
    M\to M' + \tilde\chi^0_1 + \tilde\chi^0_1\,,
\label{eq:MtoMpn1n1}
\end{equation}
related to the rare SM decays 
\begin{equation}
K\to \pi+\nu+\bar\nu\,,\footnote{Recently, the NA62 collaboration has reported, for the first time, the clear observation of this decay mode~\cite{NA62talk}, which gives a central value of BR$(K^+\to \pi^+   +  \nu  +  \bar{\nu})$ that is $1.7\sigma$ above the SM prediction. These results and their future updated ones can be used for constraining the theoretical scenarios of this work, too.}\quad  B\to K+\nu+\bar\nu\,.
\label{B-to-Knunu}
\end{equation}
Furthermore, they can lead to the decay of a baryon or to a baryon~\cite{Chamoun:2020aft,Dib:2022ppx,Domingo:2024qoj},
\textit{e.g.}
\begin{equation}
    p\to K+\tilde\chi^0_1,\, \quad B^+\to p + \tilde\chi^0_1\,.
\end{equation}

Among all these new decay modes, the decay $B^+\to K^+  + \tilde{\chi}^0_1  + \tilde{\chi}^0_1$ is relevant to the \texttt{Belle~II} measurement.
This is an R-parity-conserving decay, proceeding via supersymmetric gauge couplings, although such a light neutralino is only allowed in the RPV-SUSY.
However, the decays in Eq.~\eqref{eq:MtoMpn1n1} proceed via 1-loop diagrams and give too weak contributions~\cite{Dreiner:2009er}.
Instead, we propose the decay
\begin{equation}
    B^+\to K^+  +\overset{\scriptscriptstyle(-)}{\nu}  +\tilde{\chi}^0_1\,,
    \label{eq:signal_decay}
\end{equation}
with a massless neutralino.
This decay proceeds at tree-level in the RPV-SUSY, \textit{e.g.}~as in Fig.~\ref{fig:feynman}.
Both the neutrino and the neutralino are experimentally not observed.
Thus, this is indistinguishable from the SM decay mode, for a massless neutralino.
If the neutralino has a small but non-vanishing mass, this additional decay could lead to distortions in the decay distribution of the reconstructed invariant mass of the neutrino pair.
This could be distinguished with higher statistics.
We are not aware that this decay has been considered before.

This short note is organized as follows.
In Sect.~\ref{sec:model}, we discuss the signal processes in the RPV-SUSY.
In Sect.~\ref{sect:results}, we give details of our numerical analysis and present the numerical results.
In Sect.~\ref{sec:conslusions}, we summarize our findings and conclude.

\section{Signal processes in the RPV-SUSY}
\label{sec:model}

\begin{figure}[t]
	\centering
	\includegraphics[width=0.45\textwidth]{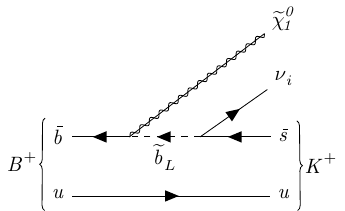}
	\caption{  \label{fig:feynman}
        One possible tree-level Feynman diagram for the decay $B^+\to K^+ + \nu_i + \tilde\chi^0_1$ via 
        a non-zero $\lambda'_{i32}$.
        Two more diagrams with an off-shell sneutrino and strange squark, respectively, are involved in the computation, albeit without being shown here.
        }
\end{figure}

The decay $B^+\to K^+ +\overset{\scriptscriptstyle(-)}{\nu_i} +\tilde\chi^0_1$  with $i=1,2,3$ can be induced via different RPV $\lambda'$ couplings.
For $\lambda'_{i32}\not=0$, it can proceed, for example, via the tree-level Feynman diagram involving a virtual scalar bottom squark, as shown in Fig.~\ref{fig:feynman}. 
There are in total three relevant tree-level diagrams. 
Analogous diagrams also exist in the case $\lam'_{i23}$ with different quark/squark chiralities.
Assuming the relevant sfermion masses to be degenerate and also much larger than the investigated mass scale, we can derive an effective four-fermion interaction by integrating out the heavy sfermions.\footnote{For such a derivation, see \textit{e.g.}~Ref.~\cite{deVries:2015mfw}.
However, Ref.~\cite{deVries:2015mfw} employs Fierz-identities with an erroneous numerical factor in the tensor current.
A corrected derivation is provided in Ref.~\cite{Gunther:2023vmz}.}
We work in a basis, where only the left-handed down-type quarks $d_{L\ell}$ are flavor eigenstates with $\ell=1,2,3$ being a generation index, whereas all other SM fields are mass eigenstates.
We express the effective interactions in terms of the mass eigenstates for the left-handed down-type quarks by taking into account the CKM-mixing $d_{L\ell}=V^{\text{CKM}} _{\ell m}d^{\mathrm{mass}}_{Lm}$ where $m\in\{1,2,3\}$, like $\ell$, is a generation index.
Here $V^{\text{CKM}}_{ij}$, $i,j\in\{1,2,3\}$ denotes the SM CKM matrix.
Under these assumptions, the decay $B^+\to K^++\nu_i+\tilde\chi^0_1$ is allowed for a non-zero $\lam^\prime_{i32}$, $\lam^\prime_{i12}$, or $\lam^\prime_{i22}$, where the latter two are CKM-suppressed. 
Similarly, $B^+\to K^++\bar\nu_i+\tilde\chi^0_1$ can be generated at tree-level via non-zero $\lam^\prime_{i23}$, and the CKM suppressed $\lam^\prime_{i13}$ and $\lam^\prime_{i33}$.

Kinematically, the decay given in Eq.~\eqref{eq:signal_decay} can only proceed if the neutralino mass satisfies
\begin{eqnarray}
   m_{\tilde\chi^0_1} < M_{B^+}- M_{K^+}\approx 4.78 \,\text{GeV}\,.
\end{eqnarray}
However, only for a very light neutralino would the kinematic distribution of the final-state neutralino-neutrino pair be indistinguishable from the SM neutrino-neutrino pair at the experimentally detectable level.
For non-universal supersymmetry breaking electroweak gaugino masses the neutralino can be arbitrarily light.
Such a light neutralino, even massless, is consistent with all laboratory, astrophysical, and cosmological data; see discussion in Sec.~\ref{sec:intro}.
In this work, we assume, for simplicity, a massless neutralino.

We compute the decay rates for the new proposed process $B^+\to K^+  +
\overset{\scriptscriptstyle(-)} {\nu}+ \tilde{\chi}^0_1\,$ using the above-mentioned effective interaction~\cite{Gunther:2023vmz}.
The relevant hadronic matrix elements $\Braket{K^+|\bar{b}\,\Gamma\, s|B^+}$ with $\Gamma\in\{\mathbbm{1},\,\gamma^\mu, \,\sigma^{\mu\nu}\}$ are parameterized by a set of three form factors $f_+(q^2)$, $f_0(q^2)$, and $f_T(q^2)$, respectively, where $q^2=(p_B-p_K)^2$ is the squared momentum transfer. $p_B$ and $p_K$ respectively denote the parent $B$-meson and daughter $K$-meson 4-momenta.
For the definition and employed parameterization of the form factors used in this work, we refer to the FLAG review~\cite{FlavourLatticeAveragingGroupFLAG:2021npn}.
We evaluate the phase space including the $q^2$-integral as described in Appendix B of Ref.~\cite{DeVries:2020jbs}.
The decay width for the signal process is then calculated to be
\begin{equation}
    \Gamma(B^+ \to K^+ + \overset{\scriptscriptstyle(-)}{\nu_i} +\tilde{\chi}^0_1)\approx \frac{0.00114\text{ GeV}^5}{\tilde{m}^4}\cdot \left\{\begin{matrix}
    |V^{\text{CKM}}_{u_jb}\lam^\prime_{ij2}|^2,&
    \\[4mm]
    |V^{\text{CKM}}_{u_js}\lam^\prime_{ij3}|^2,&
    \end{matrix}\right.
\end{equation}
where $\tilde{m}$ labels the degenerate sfermion mass.
The top entry is for the case $\nu_i$, the bottom for $\bar\nu_i$.

Before closing this section, we summarize the strongest current upper bounds on the single RPV couplings considered in this work for $\tilde{m}\!\gtrsim\! 1$ TeV, extracted from Refs.~\cite{Allanach:1999ic,Bansal:2019zak}:
\begin{equation}
    \begin{array}{cll}
        \lam'_{112} <& 0.21\displaystyle\frac{m_{\tilde{s}_R}}{1\text{ TeV}},& \lam'_{122} < 0.43\displaystyle\frac{m_{\tilde{s}_R}}{1\text{ TeV}},\\[3mm]
        \lam'_{123} <&  0.43 \displaystyle\frac{m_{\tilde{b}_R}}{1\text{ TeV}},  & \lam'_{132} < 1.04,\\[3mm]
        \lam'_{113} <& 0.21 \displaystyle\frac{m_{\tilde{b}_R}}{1\text{ TeV}},  &  \lam'_{133} < 0.0014\sqrt{\frac{m_{\tilde{b}}}{100\text{ GeV}}}, \\[3mm]
        \lam'_{212} <& 0.59\displaystyle\frac{m_{\tilde{s}_R}}{1\text{ TeV}}, &\lam'_{222} < 1.12,\\[3mm]
        \lam'_{223} <& 1.12, &\lam'_{232} < 1.04,\\
        \lam'_{213} <&  0.59 \displaystyle\frac{m_{\tilde{b}_R}}{1\text{ TeV}}, & \lam'_{233} <0.15\sqrt{\frac{m_{\tilde{b}}}{100\text{ GeV}}}\\[3mm]
        \lam'_{312} <& 0.2\displaystyle\frac{m_{\tilde{s}_R}}{1\text{ TeV}}+0.046,& \lam'_{322} < 1.12,\\[3mm]
        \lam'_{323} <& 1.12, & \lam'_{332} < 1.04,\\
        \lam'_{313} <&  0.2 \displaystyle\frac{m_{\tilde{b}_R}}{1\text{ TeV}}+0.046,&\lam'_{333} < 1.04.
    \end{array}
\label{eq:RPV_bounds}
\end{equation}

These bounds have been derived from different physical processes.
The constant bounds are due to the perturbativity requirement.
The bounds on $\lam'_{112}, \lam'_{113}, \lam'_{122},\text{ and }\lambda'_{123}$ were determined from charged-current universality~\cite{Barger:1989rk}.
The upper limits on $\lam'_{133}$ and  $\lam'_{233}$ were obtained from experimental bounds on the neutrino masses~\cite{Godbole:1992fb}.
The constraints on $\lam'_{212}$ and $\lam'_{213}$ were derived from the measurement of $R_\pi=\Gamma(\pi\to e + \nu)/\Gamma(\pi\to\mu + \nu)$~\cite{Barger:1989rk,ledroit:1998abc}.
Finally, the strongest bounds on $\lam'_{312}$ and $\lam'_{313}$ were obtained in Ref.~\cite{Bansal:2019zak} by comparing the RPV-coupling predictions with the measurements of LHC mono-lepton searches~\cite{ATLAS:2018ihk,CMS:2018fza}.

Finally, the masses of the involved sfermions, \textit{i.e.}~squarks and sneutrinos, have been bounded from below at the LHC.
For instance, Ref.~\cite{Carpenter:2020fnh} has derived the latest lower bounds on sneutrino masses from LHC mono-boson searches with a 36 fb$^{-1}$ dataset, that are below 100 GeV for scenarios with the sneutrinos as the lightest supersymmetric particle. The bounds are expected to be raised to up to about 350 GeV in the high-luminosity LHC (HL-LHC) period.
Considering simplified supersymmetric models, for the first- and second-generation (third-generation) squarks, the present bounds on their masses after the LHC Run2 are at about 2 TeV~\cite{ATLAS:2024lda} (1 TeV~\cite{Han:2016xet,Kpatcha:2021nap,Knees:2023fel,ATLAS:2024lda}), while the future HL-LHC is expected to constrain these masses at as large as around 3 TeV~\cite{Ruhr:2016xsg} (2 TeV~\cite{Baer:2023uwo,Ruhr:2016xsg}).
We are not aware of bounds on the masses of the sfermions relevant to this study, that are derived in RPV scenarios with the $\lambda'$ couplings studied here.
Since we assume degenerate sfermion masses as mentioned above, we take the minimal sfermion mass to be 1 TeV as considered in numerical results to be presented in the next section.

\section{Numerical analysis and results} 
\label{sect:results}

Here, we make a comparison between the experimental measurement of BR$^{\text{exp}}(B^+ \to K^+ + \nu +\bar{\nu})$ and the sum of the two theoretically predicted decay branching ratios BR$^{\text{SM}}(B^+\to K^++\nu+\bar{\nu})$ 
and the new decay mode BR$^{\text{RPV}}(B^+\to K^+ +\overset{\scriptscriptstyle(-)}{\nu}+\tilde{\chi}^0_1)$. 
The experimental uncertainty is known to be $\sigma_{\text{exp}}=0.7\times 10^{-5}$.
However, we must estimate the theoretical uncertainties.
The SM prediction, \textit{cf.}~Eq.~\eqref{eq:SM_prediction}, has a relative uncertainty of about 7.1\%, which combines uncertainties of the vector form factor, CKM matrix elements, as well as the full two-loop electroweak correction factors.
We make a conservative estimate for the relative uncertainty on the RPV decay  process $B^+\to K^+ +\overset{\scriptscriptstyle(-)}{\nu}+\tilde{\chi}^0_1$ to be $7.1\%\times 1.5=10.7\%$.\footnote{We have tested varying the factor 1.5 between 1.0 and 2.0 in order to check its impact on the final results of this work, and found the effect to be negligible.}
We thus choose to estimate the theoretical uncertainties of the sum of BR$^{\text{SM}}(B^+\to K^+ + \nu + \bar{\nu})$ and 
BR$^{\text{RPV}}(B^+\to K^+ + \overset{\scriptscriptstyle(-)}{\nu} + \tilde{\chi}^0_1)$ as follows:
\begin{eqnarray}
\sigma_{\text{theo}}&=&7.1\% \times \text{BR}^{\text{SM}}(B^+\to K^+ + \nu +\bar{\nu}) \nonumber\\
   &&+ \, 10.7\%\times \text{BR}^{\text{RPV}}(B^+\to K^+ +\overset{\scriptscriptstyle(-)}{\nu}  +  \tilde{\chi}^0_1).\,\,\,\,\,\quad
\end{eqnarray}
Finally, we take the quadrature of both the theoretical and experimental uncertainties to obtain the 
total uncertainty:
\begin{eqnarray}
\sigma=\sqrt{\sigma_{\text{theo}}^2 + \sigma_{\text{exp}}^2}\;.
\end{eqnarray}
We proceed to define the significance as
\begin{eqnarray}
    S &=&\Big|\,\text{BR}^{\text{RPV}}(B^+\to K^+ + \overset{\scriptscriptstyle(-)}{\nu} + \tilde{\chi}^0_1) \nonumber\\
    &&\;+ \text{BR}^{\text{SM}}(B^+\to K^+ + \nu + \bar{\nu}) \nonumber\\ 
    &&\;-\, \text{BR}^{\text{exp}}(B^+ \to K^+  +\nu + \bar{\nu})\,\Big|/\sigma\,.
    \label{eq:significance}
\end{eqnarray}

\begin{figure}[t]
	\centering
	\includegraphics[width=0.5\textwidth]{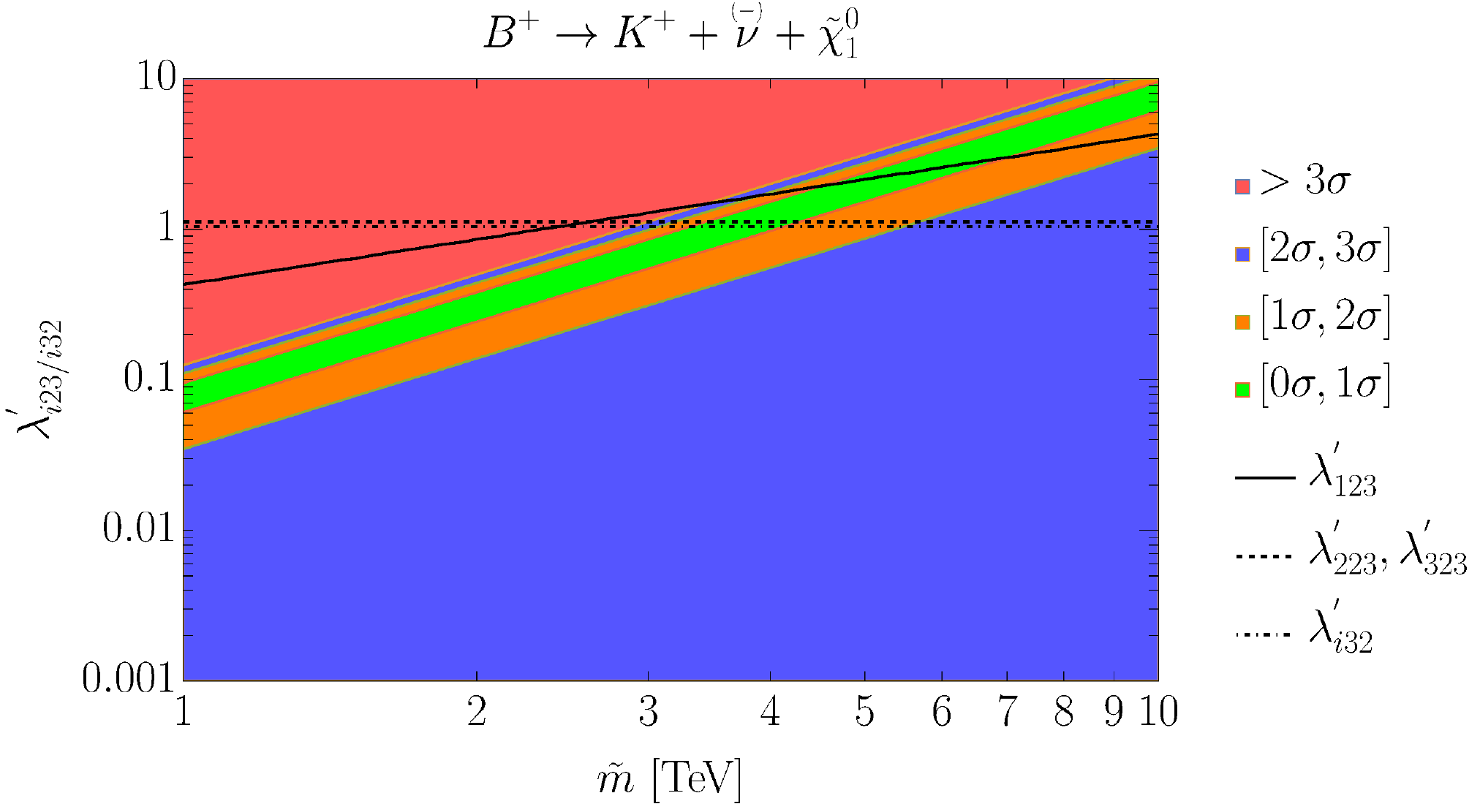}
	\caption{Constraints from the new \texttt{Belle~II} measurement on RPV couplings $\lam'_{i23}$ and $\lam'_{i32}$ as a function of the degenerate sfermion masses.
 Here, we assume the new supersymmetric decay $B\to K+\overset{(-)}{\nu}+\tilde\chi^0_1$ explains the  \texttt{Belle~II} anomaly.
 Various significance regions are shown with different colors, and the current upper bounds on the RPV couplings are included.}
 \label{fig:results1}
\end{figure}

\begin{figure}[t]
	\centering
	\includegraphics[width=0.5\textwidth]{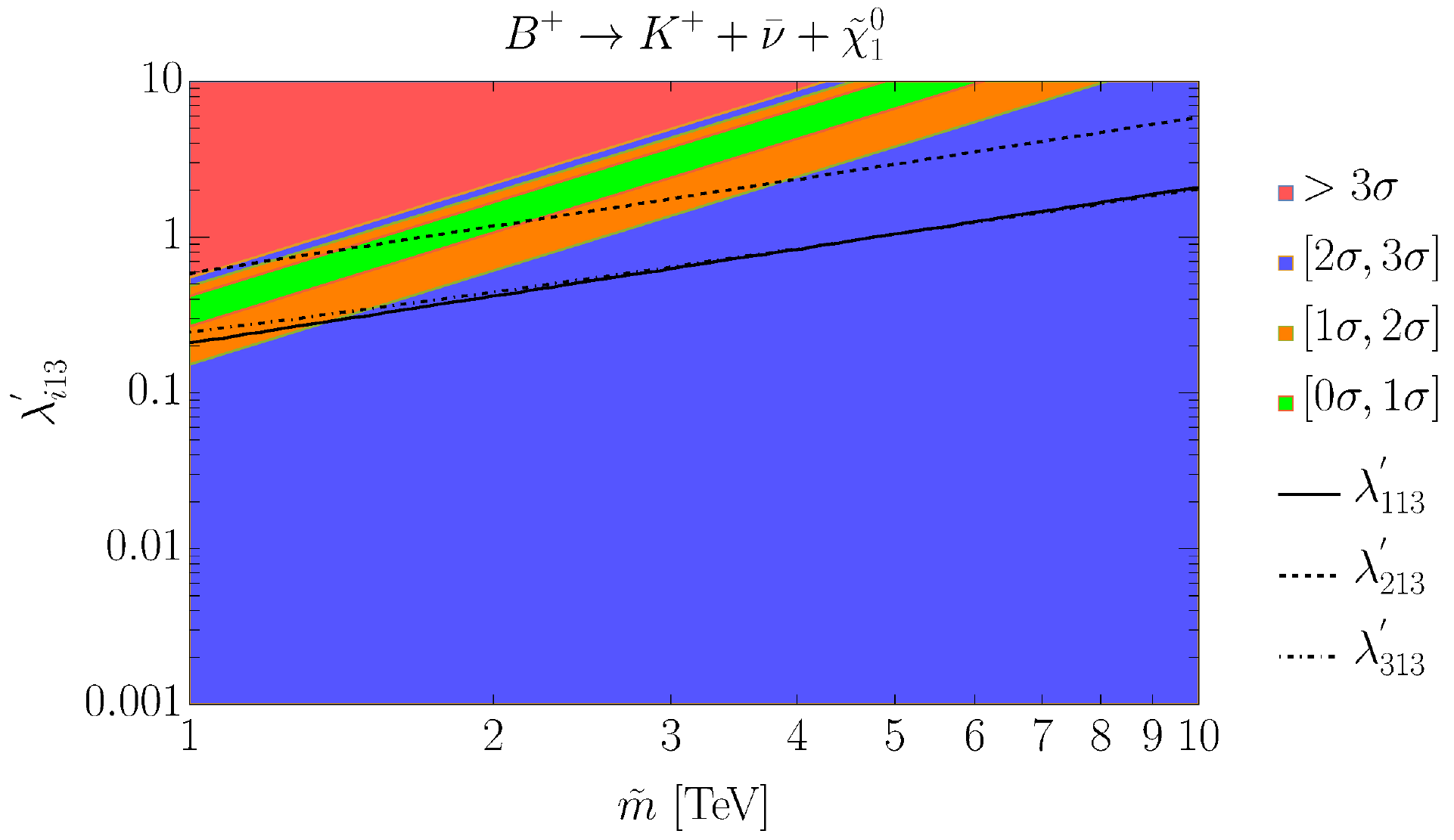}
	\caption{The same as Fig.~\ref{fig:results1} but for $\lam'_{i13}$.}
        \label{fig:results2}
\end{figure}

While various RPV couplings can mediate the signal process studied in this work, we choose not to show results for the couplings $\lam'_{i33}$, $\lam'_{i12}$ and $\lam'_{i22}$.
In these cases we find that at most only a tiny part of the parameter space could simultaneously accommodate the new measurement and satisfy the current bounds on the corresponding RPV couplings.
Thus, we present the numerical results in Fig.~\ref{fig:results1} and Fig.~\ref{fig:results2} in the $(\tilde{m},\lam')$ plane, for the RPV couplings $\lam'_{i23/i32}$ and $\lam'_{i13}$, respectively.
The results for $\lam'_{i23}$ and $\lam'_{i32}$ are almost identical and therefore overlaid in a single plot.
The range of the sfermion masses is taken to be $[1\text{ TeV}, 10\text{ TeV}]$.
The parameter space is divided into different colored regions according to the significance level as computed from Eq.~\eqref{eq:significance}: 
green for $[0, 1\,\sigma]$, orange for $[1\,\sigma, 2\,\sigma]$, blue for $[2\,\sigma, 3\,\sigma]$, and red for $[>3\,\sigma]$.
Furthermore, we show together the current upper bounds on the involved RPV couplings with black lines of distinct line styles; see Eq.~\eqref{eq:RPV_bounds}.

We find the scenarios with $\lam'_{i23}$ or $\lam'_{i32}$ the most promising, as shown in Fig.~\ref{fig:results1}.
Large parts of the parameter space can accommodate the latest \texttt{Belle~II} measurement within a $2\,\sigma$ significance.
The current RPV bounds can exclude only a small part of the parameter space with large coupling and large sfermion masses.
This is especially the case for $\lam'_{123}$.
The results shown in Fig.~\ref{fig:results2} for the couplings $\lambda'_{i13}$ are weaker, mainly as a result of the CKM suppression.
They show that only for $\lam'_{213}$ the majority of the parameter region that explains the anomaly within $2\sigma$ is still allowed, while with the first- and third-generation leptons the present bounds almost completely rule out the $2\,\sigma$ regions.

\section{Summary and Conclusions}\label{sec:conslusions}

The recent experimental evidence for the rare decay process $B^+\to K^++\nu+\bar{\nu}$ at $3.5\,\sigma$ obtained by the \texttt{Belle~II} experiment corresponds to the decay branching ratio $2.3 \times 10^{-5}$, which is above the SM prediction by $2.7\,\sigma$.\footnote{In Ref.~\cite{Belle-II:2023esi}, a new average value of BR$(B^+\to K^++\nu+\bar{\nu})$ at $(1.3\pm0.4)\times 10^{-5}$, obtained by taking into account previous measurement results from Belle~\cite{Belle:2013tnz,Belle:2017oht} and BaBar~\cite{BaBar:2010oqg,BaBar:2013npw} is also given. It is above the SM prediction by just about two standard deviations. As a result, if we use this value for the analysis performed here,  we will derive only upper bounds on the considered RPV scenarios at $2\sigma$ significance, with no lower bounds, and these upper bounds will be slightly stronger than those presented in this work.}
Here, we have proposed to explain this enhancement in the branching ratio, by considering a massless bino in RPV-SUSY, which is currently allowed by all experimental and observational constraints.
With a single RPV coupling $\lambda'_{i23/i32}$, $\lambda'_{i13/i33}$, or $\lambda'_{i12/i22}$, turned on, the $B^+$ meson can decay to the $K^+\overset{\scriptscriptstyle(-)}{\nu}\tilde{\chi}^0_1$ final state, mimicking the SM signature, the supposedly observed $K^+ \nu \bar{\nu}$ final state.

We have computed the decay width $\Gamma(B^+\to K^+ + \overset{\scriptscriptstyle(-)}{\nu} + \tilde\chi^0_1)$ in the RPV-SUSY for the mentioned couplings.
We have estimated the significance levels for accommodating the new measurement with the RPV supersymmetric model, taking into account both theoretical and experimental uncertainties.
We have also considered the current bounds on the considered RPV couplings.
Our numerical findings displayed in Fig.~\ref{fig:results1} show that $\lam'_{i23}$ and $\lam'_{i32}$ have almost identical results, and large parts of the parameter space spanned by $\lam'_{i23/i32}$ and the degenerate sfermion mass $\tilde{m}$ can explain the latest \texttt{Belle~II} measurement within the $2\,\sigma$ significance level, while still obeying the current bounds on these RPV couplings.
On the other hand, for $\lam'_{i13/i33}$ and $\lam'_{i12/i22}$, the RPV contributions of $B^+ \to K^+ + \overset{\scriptscriptstyle(-)}{\nu} +\tilde{\chi}^0_1$ are CKM-suppressed, rendering, in almost all the cases, only a tiny proportion or even none of the parameter space accommodating the new measurement and satisfying the current bounds on the RPV couplings simultaneously.
Therefore, for these scenarios, we choose to present the numerical results for the coupling $\lambda'_{i13}$ only, given in Fig.~\ref{fig:results2}, where we find that for $\lambda'_{213}$ a large part of the parameter region explaining the observed enhancement in BR$(B^+\to K^+ + \nu  +  \bar{\nu})$ within $2\,\sigma$ significance is still allowed.

Before concluding, we mention that current upper bounds on BR$(B\to K^*+\nu+\bar{\nu})$~\cite{Belle:2017oht,ParticleDataGroup:2024cfk} can, in principle, also be used for deriving bounds on the $\lambda'$ couplings as functions of sfermion masses.
These bounds are measured to be of the same order of magnitude as the measured value of BR$(B^+\to K^+ + \nu + \bar{\nu})$ at Belle~II, with the allowed value of the branching ratios contributed from new physics also being of the same order of magnitude~\cite{He:2022ljo}.
Since our scenarios with a light bino in the RPV-SUSY have both scalar and tensor (pseudoscalar and tensor) operators contributing to the $B\to K$ ($B\to K^*$) transitions (see Appendix A.3 and Appendix C of Ref.~\cite{Gunther:2023vmz}), and since the differences in the form factors, kaon masses, as well as the phase-space suppression factors, should have only minor impact on the decay rates, we expect that using the measured upper bounds on BR$(B\to K^* + \nu + \bar{\nu})$ would lead to constraints on the RPV couplings similar to those derived here.
Therefore, we choose not to compute these bounds in detail in this work.

Finally, we comment that future precision measurements of low-energy observables and collider searches can either confirm this scenario with a massless bino or place stronger bounds on the relevant RPV couplings.

\medskip

\bigskip
\centerline{\bf Acknowledgements}

\bigskip
We thank Jordy de Vries for inspiring discussions, and Carlos Wagner for frequent discussions of his related stimulating hypothesis.
HKD acknowledges partial financial support by the Deutsche Forschungsgemeinschaft
(DFG, German Research Foundation) through the funds provided to the Sino-German
Collaborative Research Center TRR110 ``Symmetries and the Emergence of Structure in
QCD'' (DFG Project ID 196253076 — TRR 110).

\bigskip

\bibliography{refs}

\begin{thebibliography}{10}

\bibitem{Belle-II:2023esi}
Belle-II, I.~Adachi {\em et~al.},
\newblock Phys. Rev. D {\bf 109}, 112006 (2024), 2311.14647.

\bibitem{Parrott:2022zte}
HPQCD, W.~G. Parrott, C.~Bouchard, and C.~T.~H. Davies,
\newblock Phys. Rev. D {\bf 107}, 014511 (2023), 2207.13371,
\newblock [Erratum: Phys.Rev.D 107, 119903 (2023)].

\bibitem{Becirevic:2023aov}
D.~Be\v{c}irevi\'c, G.~Piazza, and O.~Sumensari,
\newblock Eur. Phys. J. C {\bf 83}, 252 (2023), 2301.06990.

\bibitem{Asadi:2023ucx}
P.~Asadi, A.~Bhattacharya, K.~Fraser, S.~Homiller, and A.~Parikh,
\newblock JHEP {\bf 10}, 069 (2023), 2308.01340.

\bibitem{Athron:2023hmz}
P.~Athron, R.~Martinez, and C.~Sierra,
\newblock JHEP {\bf 02}, 121 (2024), 2308.13426.

\bibitem{Bause:2023mfe}
R.~Bause, H.~Gisbert, and G.~Hiller,
\newblock Phys. Rev. D {\bf 109}, 015006 (2024), 2309.00075.

\bibitem{Allwicher:2023xba}
L.~Allwicher, D.~Becirevic, G.~Piazza, S.~Rosauro-Alcaraz, and O.~Sumensari,
\newblock Phys. Lett. B {\bf 848}, 138411 (2024), 2309.02246.

\bibitem{Felkl:2023ayn}
T.~Felkl, A.~Giri, R.~Mohanta, and M.~A. Schmidt,
\newblock Eur. Phys. J. C {\bf 83}, 1135 (2023), 2309.02940.

\bibitem{He:2023bnk}
X.-G. He, X.-D. Ma, and G.~Valencia,
\newblock Phys. Rev. D {\bf 109}, 075019 (2024), 2309.12741.

\bibitem{Chen:2023wpb}
C.-H. Chen and C.-W. Chiang,
\newblock Phys. Rev. D {\bf 109}, 075004 (2024), 2309.12904.

\bibitem{Berezhnoy:2023rxx}
A.~Berezhnoy and D.~Melikhov,
\newblock EPL {\bf 145}, 14001 (2024), 2309.17191.

\bibitem{Alonso-Alvarez:2023mgc}
G.~Alonso-\'Alvarez and M.~Escudero~Abenza,
\newblock Eur. Phys. J. C {\bf 84}, 553 (2024), 2310.13043.

\bibitem{DaRold:2023hmx}
L.~Da~Rold,
\newblock JHEP {\bf 06}, 152 (2024), 2311.04062.

\bibitem{Altmannshofer:2023hkn}
W.~Altmannshofer, A.~Crivellin, H.~Haigh, G.~Inguglia, and J.~Martin~Camalich,
\newblock Phys. Rev. D {\bf 109}, 075008 (2024), 2311.14629.

\bibitem{Fridell:2023ssf}
K.~Fridell, M.~Ghosh, T.~Okui, and K.~Tobioka,
\newblock Phys. Rev. D {\bf 109}, 115006 (2024), 2312.12507.

\bibitem{Ho:2024cwk}
S.-Y. Ho, J.~Kim, and P.~Ko,
\newblock (2024), 2401.10112.

\bibitem{Hou:2024vyw}
B.-F. Hou, X.-Q. Li, M.~Shen, Y.-D. Yang, and X.-B. Yuan,
\newblock JHEP {\bf 06}, 172 (2024), 2402.19208.

\bibitem{Chen:2024cll}
C.-H. Chen and C.-W. Chiang,
\newblock (2024), 2403.02897.

\bibitem{Bolton:2024egx}
P.~D. Bolton, S.~Fajfer, J.~F. Kamenik, and M.~Novoa-Brunet,
\newblock Phys. Rev. D {\bf 110}, 055001 (2024), 2403.13887.

\bibitem{Rosauro-Alcaraz:2024mvx}
S.~Rosauro-Alcaraz and L.~P.~S. Leal,
\newblock Eur. Phys. J. C {\bf 84}, 795 (2024), 2404.17440.

\bibitem{Buras:2024ewl}
A.~J. Buras, J.~Harz, and M.~A. Mojahed,
\newblock JHEP {\bf 10}, 087 (2024), 2405.06742.

\bibitem{Dreiner:1991pe}
H.~K. Dreiner and G.~G. Ross,
\newblock Nucl. Phys. B {\bf 365}, 597 (1991).

\bibitem{Allanach:2003eb}
B.~C. Allanach, A.~Dedes, and H.~K. Dreiner,
\newblock Phys. Rev. D {\bf 69}, 115002 (2004), hep-ph/0309196,
\newblock [Erratum: Phys.Rev.D 72, 079902 (2005)].

\bibitem{Barbier:2004ez}
R.~Barbier {\em et~al.},
\newblock Phys. Rept. {\bf 420}, 1 (2005), hep-ph/0406039.

\bibitem{Choudhury:1999tn}
D.~Choudhury, H.~K. Dreiner, P.~Richardson, and S.~Sarkar,
\newblock Phys. Rev. D {\bf 61}, 095009 (2000), hep-ph/9911365.

\bibitem{Dreiner:2009er}
H.~K. Dreiner {\em et~al.},
\newblock Phys. Rev. D {\bf 80}, 035018 (2009), 0905.2051.

\bibitem{Dreiner:2009ic}
H.~K. Dreiner {\em et~al.},
\newblock Eur. Phys. J. C {\bf 62}, 547 (2009), 0901.3485.

\bibitem{Gogoladze:2002xp}
I.~Gogoladze, J.~D. Lykken, C.~Macesanu, and S.~Nandi,
\newblock Phys. Rev. D {\bf 68}, 073004 (2003), hep-ph/0211391.

\bibitem{Dreiner:2022swd}
H.~K. Dreiner, D.~K\"ohler, S.~Nangia, and Z.~S. Wang,
\newblock JHEP {\bf 02}, 120 (2023), 2207.05100.

\bibitem{Dreiner:2003wh}
H.~K. Dreiner, C.~Hanhart, U.~Langenfeld, and D.~R. Phillips,
\newblock Phys. Rev. D {\bf 68}, 055004 (2003), hep-ph/0304289.

\bibitem{Dreiner:2011fp}
H.~K. Dreiner, M.~Hanussek, J.~S. Kim, and S.~Sarkar,
\newblock Phys. Rev. D {\bf 85}, 065027 (2012), 1111.5715.

\bibitem{Dreiner:2013tja}
H.~K. Dreiner, J.-F. Fortin, J.~Isern, and L.~Ubaldi,
\newblock Phys. Rev. D {\bf 88}, 043517 (2013), 1303.7232.

\bibitem{Dercks:2018wum}
D.~Dercks, H.~K. Dreiner, M.~Hirsch, and Z.~S. Wang,
\newblock Phys. Rev. D {\bf 99}, 055020 (2019), 1811.01995.

\bibitem{Dreiner:2023gir}
H.~K. Dreiner, D.~K\"ohler, S.~Nangia, M.~Sch\"urmann, and Z.~S. Wang,
\newblock JHEP {\bf 08}, 058 (2023), 2306.14700.

\bibitem{Barger:1989rk}
V.~D. Barger, G.~F. Giudice, and T.~Han,
\newblock Phys. Rev. D {\bf 40}, 2987 (1989).

\bibitem{Allanach:1999ic}
B.~C. Allanach, A.~Dedes, and H.~K. Dreiner,
\newblock Phys. Rev. D {\bf 60}, 075014 (1999), hep-ph/9906209.

\bibitem{Dedes:2001zia}
A.~Dedes, H.~K. Dreiner, and P.~Richardson,
\newblock Phys. Rev. D {\bf 65}, 015001 (2001), hep-ph/0106199.

\bibitem{deVries:2015mfw}
J.~de~Vries, H.~K. Dreiner, and D.~Schmeier,
\newblock Phys. Rev. D {\bf 94}, 035006 (2016), 1511.07436.

\bibitem{Dercks:2018eua}
D.~Dercks, J.~De~Vries, H.~K. Dreiner, and Z.~S. Wang,
\newblock Phys. Rev. D {\bf 99}, 055039 (2019), 1810.03617.

\bibitem{Dreiner:2020qbi}
H.~K. Dreiner, J.~Y. G\"unther, and Z.~S. Wang,
\newblock Phys. Rev. D {\bf 103}, 075013 (2021), 2008.07539.

\bibitem{Adhikari:1994iv}
R.~Adhikari and B.~Mukhopadhyaya,
\newblock Phys. Lett. B {\bf 353}, 228 (1995), hep-ph/9411208.

\bibitem{NA62talk}
NA62, J.~Swallow,
\newblock {New measurement of the $K^+ \to \pi^+ \nu \bar\nu$ decay by the NA62
  Experiment: \url{https://indico.cern.ch/event/1447422/}},
\newblock 2024.

\bibitem{Chamoun:2020aft}
N.~Chamoun, F.~Domingo, and H.~K. Dreiner,
\newblock Phys. Rev. D {\bf 104}, 015020 (2021), 2012.11623.

\bibitem{Dib:2022ppx}
C.~O. Dib {\em et~al.},
\newblock JHEP {\bf 02}, 224 (2023), 2208.06421.

\bibitem{Domingo:2024qoj}
F.~Domingo, H.~K. Dreiner, D.~K\"ohler, S.~Nangia, and A.~Shah,
\newblock JHEP {\bf 05}, 258 (2024), 2403.18502.

\bibitem{Gunther:2023vmz}
J.~Y. G\"unther, J.~de~Vries, H.~K. Dreiner, Z.~S. Wang, and G.~Zhou,
\newblock JHEP {\bf 01}, 108 (2024), 2310.12392.

\bibitem{FlavourLatticeAveragingGroupFLAG:2021npn}
Flavour Lattice Averaging Group (FLAG), Y.~Aoki {\em et~al.},
\newblock Eur. Phys. J. C {\bf 82}, 869 (2022), 2111.09849.

\bibitem{DeVries:2020jbs}
J.~De~Vries, H.~K. Dreiner, J.~Y. G\"unther, Z.~S. Wang, and G.~Zhou,
\newblock JHEP {\bf 03}, 148 (2021), 2010.07305.

\bibitem{Bansal:2019zak}
S.~Bansal, A.~Delgado, C.~Kolda, and M.~Quiros,
\newblock Phys. Rev. D {\bf 100}, 093005 (2019), 1906.01063.

\bibitem{Godbole:1992fb}
R.~M. Godbole, P.~Roy, and X.~Tata,
\newblock Nucl. Phys. B {\bf 401}, 67 (1993), hep-ph/9209251.

\bibitem{ledroit:1998abc}
F.~Ledroit and G.~Sajot,
\newblock GDR-S-008 (ISN, Grenoble, 1998)  (1998).

\bibitem{ATLAS:2018ihk}
ATLAS, M.~Aaboud {\em et~al.},
\newblock Phys. Rev. Lett. {\bf 120}, 161802 (2018), 1801.06992.

\bibitem{CMS:2018fza}
CMS, A.~M. Sirunyan {\em et~al.},
\newblock Phys. Lett. B {\bf 792}, 107 (2019), 1807.11421.

\bibitem{Carpenter:2020fnh}
L.~M. Carpenter, H.~B. Gilmer, and J.~Kawamura,
\newblock Phys. Rev. D {\bf 103}, 095014 (2021), 2007.10360.

\bibitem{ATLAS:2024lda}
ATLAS, G.~Aad {\em et~al.},
\newblock (2024), 2403.02455.

\bibitem{Han:2016xet}
C.~Han, J.~Ren, L.~Wu, J.~M. Yang, and M.~Zhang,
\newblock Eur. Phys. J. C {\bf 77}, 93 (2017), 1609.02361.

\bibitem{Kpatcha:2021nap}
E.~Kpatcha {\em et~al.},
\newblock Eur. Phys. J. C {\bf 82}, 261 (2022), 2111.13212.

\bibitem{Knees:2023fel}
P.~Knees, E.~Kpatcha, I.~n. Lara, D.~E. L\'opez-Fogliani, and C.~Mu\~noz,
\newblock Eur. Phys. J. C {\bf 84}, 104 (2024), 2309.06456.

\bibitem{Ruhr:2016xsg}
ATLAS, F.~R\"uhr,
\newblock Nucl. Part. Phys. Proc. {\bf 273-275}, 625 (2016).

\bibitem{Baer:2023uwo}
H.~Baer, V.~Barger, J.~Dutta, D.~Sengupta, and K.~Zhang,
\newblock Phys. Rev. D {\bf 108}, 075027 (2023), 2307.08067.

\bibitem{Belle:2013tnz}
Belle, O.~Lutz {\em et~al.},
\newblock Phys. Rev. D {\bf 87}, 111103 (2013), 1303.3719.

\bibitem{Belle:2017oht}
Belle, J.~Grygier {\em et~al.},
\newblock Phys. Rev. D {\bf 96}, 091101 (2017), 1702.03224,
\newblock [Addendum: Phys.Rev.D 97, 099902 (2018)].

\bibitem{BaBar:2010oqg}
BaBar, P.~del Amo~Sanchez {\em et~al.},
\newblock Phys. Rev. D {\bf 82}, 112002 (2010), 1009.1529.

\bibitem{BaBar:2013npw}
BaBar, J.~P. Lees {\em et~al.},
\newblock Phys. Rev. D {\bf 87}, 112005 (2013), 1303.7465.

\bibitem{ParticleDataGroup:2024cfk}
Particle Data Group, S.~Navas {\em et~al.},
\newblock Phys. Rev. D {\bf 110}, 030001 (2024).

\bibitem{He:2022ljo}
X.-G. He, X.-D. Ma, and G.~Valencia,
\newblock JHEP {\bf 03}, 037 (2023), 2209.05223.

\end{thebibliography}
\bibliographystyle{h-physrev}

\end{document}